\documentclass[journal=jacsat,manuscript=article]{achemso}

\usepackage[version=3]{mhchem} 

\usepackage{color}

\usepackage{graphicx}
\usepackage[space]{grffile}
\usepackage{latexsym}
\usepackage{amsfonts,amsmath,amssymb}
\usepackage{url}
\usepackage[utf8]{inputenc}
\usepackage{fancyref}
\usepackage{hyperref}
\hypersetup{colorlinks=false,pdfborder={0 0 0},}
\usepackage{textcomp}
\usepackage{longtable}
\usepackage{multirow,booktabs}

\author{Kainen L. Utt}
\affiliation{Department of Physics. University of Arkansas, Fayetteville, AR 72701}
\author{Pablo Rivero}
\affiliation{Department of Physics. University of Arkansas, Fayetteville, AR 72701}
\author{Mehrshad Mehboudi}
\affiliation{Department of Physics. University of Arkansas, Fayetteville, AR 72701}
\author{Edmund O. Harriss}
\affiliation{Department of Mathematical Sciences. University of Arkansas, Fayetteville, AR 72701}
\author{Mario F. Borunda}
\affiliation{Department of Physics. Oklahoma State University, Stillwater OK 74078}
\author{Alejandro A. Pacheco SanJuan}
\affiliation{Departamento de Ingenier{\'\i}a Mec{\'a}nica. Universidad del Norte. Barranquilla, Colombia}
\email{apacheco@uninorte.co}
\author{Salvador Barraza-Lopez}
\affiliation{Department of Physics. University of Arkansas, Fayetteville, AR 72701}
\email{sbarraza@uark.edu}

\title{Intrinsic defects, fluctuations of the local shape,\\and the photo-oxidation of black phosphorus}

\begin{document}

\begin{tocentry}
\begin{center}
\includegraphics{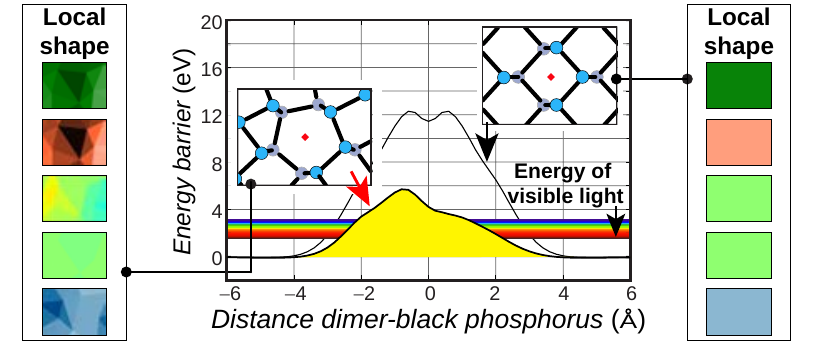}
\end{center}
{\bf Intrinsic defects, fluctuations of the local shape, and the photo-oxidation of black phosphorus}\\
Authors: Utt, Kainen; Rivero, Pablo; Mehboudi, Mehrshad; Harriss, Edmund; Borunda, Mario; Pacheco SanJuan, Alejandro; Barraza-Lopez, Salvador.\\

{\bf Synopsis statement:} Intrinsic defects induce fluctuations of a local shape. The energy barrier for oxygen dimers to pierce black phosphorus reduces its magnitude there within the realm of light-induced excitations. Defects dissociate oxygen dimers too.

\end{tocentry}

\begin{abstract}
Black phosphorus is a monoatomic semiconducting layered material that degrades exothermically in the presence of light and ambient contaminants. Its degradation dynamics remain largely unknown. Even before degradation, local-probe studies indicate non-negligible local curvature --through a non-constant height distribution-- due to the unavoidable presence of intrinsic defects. We establish that these intrinsic defects are photo-oxidation sites because they lower the chemisorption barrier of ideal black phosphorus ($> 10$ eV and out of visible-range light excitations) right into the visible and ultra-violet range (1.6 to 6.8 eV), thus enabling photo-induced oxidation and dissociation of oxygen dimers. A full characterization of the material's shape and of its electronic properties at the early stages of the oxidation process is presented as well. This study thus provides fundamental insights into the degradation dynamics of this novel layered material.
\end{abstract}

\section{Introduction}
The understanding and prevention of oxidation processes underpins many successful technologies such as the galvanization of metals, steel manufacturing, and the controlled oxidation of silicon wafers, to mention a few examples. Furthermore, the exposure to ambient illumination is behind the oxidation and subsequent degradation (also known as {\em weathering}) of commercial polymers. Two-dimensional atomic materials\cite{2D1,2D2} like graphene,\cite{gr1,gr2} hexagonal boron nitride,\cite{hbn1,hbn2} and transition-metal dichalcogenides\cite{TMD1,TMD2,TMD3,TMD4,TMD5,TMD6,TMD7} are chemically stable when exposed to light under standard ambient conditions. On the other hand, black phosphorus (BP)\cite{BP1,BP2,BP3,BP4,BP5,BP6}   has a remarkable puckered structure that anticipates its unique chemical properties\cite{castellanos,deg0,deg1,deg2,Martel}. Uncapped BP transistors exposed to light and to ambient conditions break down after a few hours even though the stack contains a large number of monolayers, thus implying a complete material degradation that originates from random locations at the exposed layer. Capping by an AlO$_x$ overlayer\cite{deg2,deg3} or by hexagonal boron nitride (hBN)\cite{deg4,deg5,deg6,deg7} helps to prevent this acute degradation process. Understanding the oxidation of BP is a pressing and relevant problem with deep consequences for the science and engineering of this layered material and we establish that, among other mechanisms, intrinsic defects can initiate the light-induced degradation of BP: one must lower the reactivity of structural defects to delay its photo-oxidation.

A non-constant height profile is apparent in BP even prior to degradation,\cite{deg1,deg2,deg4} implying the existence of intrinsic structural defects\cite{Yakobson,Liang,us5,Liu} and a non-zero local curvature through the surface. The degradation of uncapped samples leads to a visible increase of these height variations, and hence to an increased local curvature \cite{castellanos,deg0,deg1,deg2} that manifests as pits, bubbles, and bulges.  The properties of membrane-like materials can be tuned by changes in their local shape\cite{Nelson,vitelli,EL,m1,m2,Kamien3,ngnm,Kamien1,Kamien2,vozmediano92,Haddon,vozmediano}, and it will be shown that the propensity of BP to degrade is ultimately linked to its local geometry.

Chemisorption barriers for oxygen dimers (O$_2$) on ideal BP are larger than ten electron-Volts (two-hundred and fifty kcal/mol, or a thousand kJ/mol) and cannot be accessed through optical excitations in the visible spectrum (390 to 780 nm, or 1.6 to 3.2 eV). We realize, nevertheless, that these barriers are largely reduced at intrinsic defects and take on typical values for photoinduced chemical reactions.

Indeed, photoexcitations within 3.7 to 5.0 eV lead to the photochemistry of adenine and aminopurine,\cite{serrano} setting an energy scale for reactions that could be activated by light. Light introduces sufficient energy to break or reorganize most covalent bonds and enables reactions that are otherwise thermodynamically forbidden, given that activation barriers of the order of a few eV ($>$10,000 Kelvin) are overcome.

This article represents a departure from other theoretical works\cite{oxygenplanar1,oxygenplanar2,oxygenplanar3,oxygenplanar4,oxygenplanar5,oxygenplanar6} that study the degradation of BP as we focus on the role played by intrinsic defects on this material's chemistry. This mechanistic study begins with a structural and electronic ground state consisting of a planar or conical BP monolayer and non-interacting O$_2$ molecules. We create excited atomistic (hence electronic) structures by bringing O$_2$ molecules through the BP monolayer to gain insight into the energetics involved. We then pursue new equilibrium oxidized structures with {\em ab initio} atomistic optimizations in which the O$_2$ molecules have been chemisorbed onto the BP monolayer, and provide a thorough geometrical and electronic characterization of these systems towards the end of the manuscript.  Although our present focus is on the oxidation by O$_2$ for a self-contained discussion, similar studies can be performed to understand the degradation of BP from water molecules and other ambient contaminants. An important component of this study is a tool to analyze the shape of two-dimensional materials that we have developed \cite{us1,us2,us3,us4,us5}.

\section{Results and discussion}
\subsection{The energy for O$_2$ to attach to black phosphorus depends on the local shape}\label{sec:sec2}

We first investigate how much energy is needed for O$_2$ to pierce through a BP monolayer. Two finite (aperiodic) structures are considered for this purpose: a planar one containing about 600 atoms, and a conical one containing about 500 atoms\cite{us5} (see Methods).

The local atomistic geometry is established from five local quantities that indicate how the material elongates/compresses (the trace and determinant of the metric Tr(g) and Det(g), where g is the metric tensor) and/or curves (through the mean curvature $H$, and the gaussian curvature $K$) along two orthogonal directions\cite{us3,us4,us5}, and how the local thickness $\tau$ changes with respect to its value $\tau_0$ for an ideal crystalline BP monolayer. BP is a buckled material, and $\tau_0$ is the distance among the two sublayers $s_1$ and $s_2$ that make up a monolayer.  Extensive details of the discrete geometry employed to characterize the shape of BP are given in the Methods section \cite{us5}.

\begin{figure}[tb]
\includegraphics[width=1.0\textwidth]{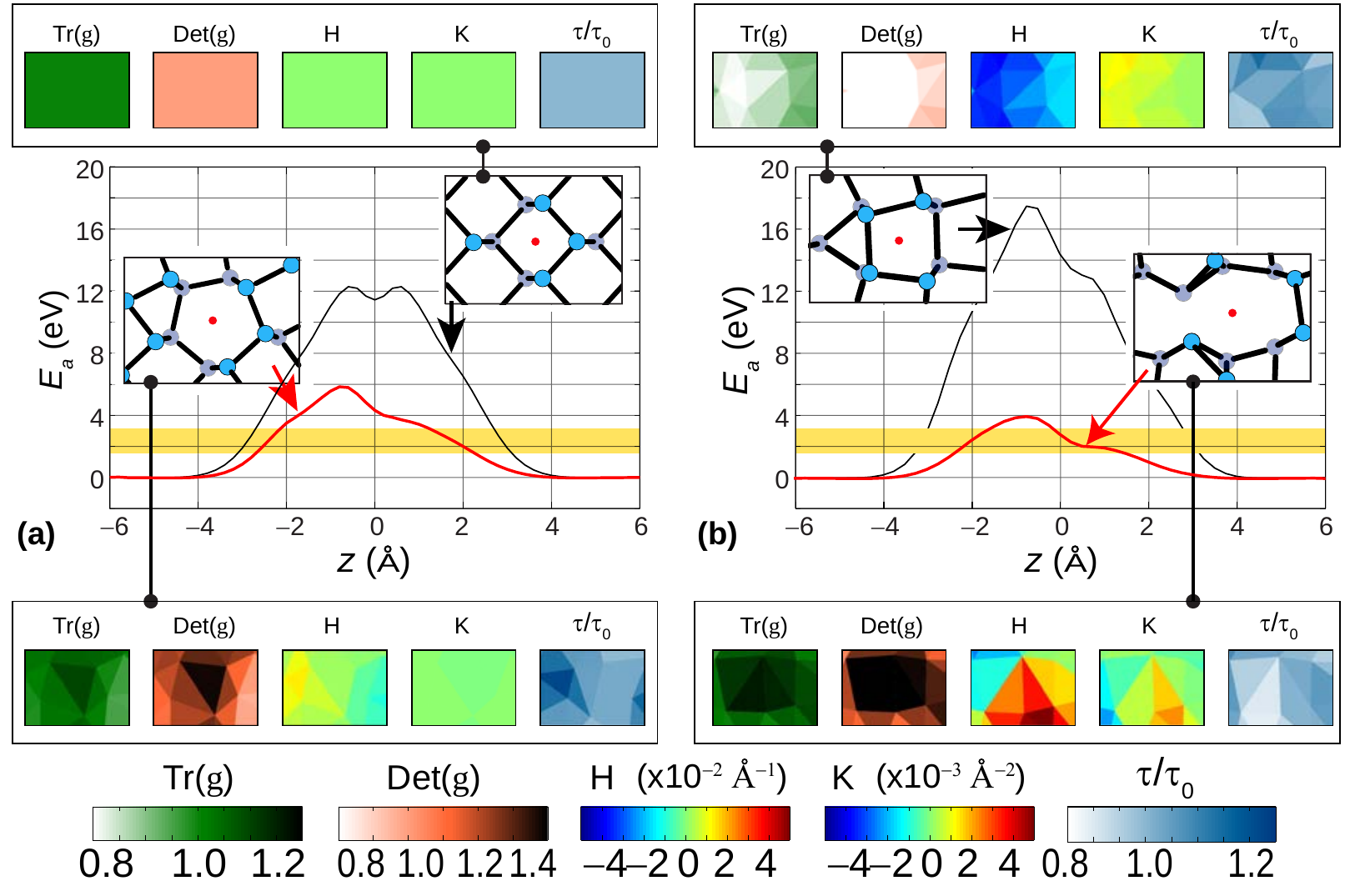}
\caption{The energy barrier $E_a$ for O$_2$ to oxidize BP depends on the local shape, and the smallest barriers on (a) planar or (b) conical structures occur at intrinsic defects. Atomistic structures and the local geometry at one sublayer ($s_1$) are shown as insets. The darker/lighter color seen on Tr(g) and Det(g) implies larger/smaller interatomic distances at these defects. The low activation barriers facilitated by intrinsic defects are within the visible electromagnetic spectrum, which is highlighted by the yellow rectangles within the 1.6--3.2 eV energy range.}
\label{fig:fig1}
\end{figure}

Activation barriers $E_a(z)$ for the oxidation of BP  that take full consideration of the spin-polarization of the oxygen dimers are estimated by bringing a vertically-oriented O$_2$ molecule --with its center of mass located at a height $z$ along the local normal-- into proximity of the BP structure, and crossing BP at the largest possible distance among O and P atoms (see reds dots on the structural insets on Figure \ref{fig:fig1}). These activation barriers are computed with {\em ab initio} calculations without performing any structural optimization (see Methods).

As seen in Figure \ref{fig:fig1}a, an oxygen dimer requires about 12 eV to rest at the middle of an ideal BP monolayer that has the uniform geometry shown in the upper inset ($Tr(g)=1$, $Det(g)=1$, $H=0$, $K=0$, and $\tau/\tau_0=1$) in Figure \ref{fig:fig1}a. The energy barrier is symmetric around $z=0$, where the dimer's center of mass coincides with the center of mass of the two buckled BP sublayers.\cite{us5} Such a high energy barrier makes the chemisorption of O$_2$ onto ideal BP rather unlikely.

However, the maximum magnitude of $E_a$ decreases significantly within intrinsic defects: for instance, the line defect with Burgers vector (0,1) \cite{Nelson} induces a dislocation line containing pentagon/heptagon pairs. \cite{Yakobson} The energy range for visible light is within 1.6 and 3.2 eV, corresponding to wavelengths within 390 to 780 nm (yellow rectangles in Figure \ref{fig:fig1}). The maximum value taken by $E_a$ as an O$_2$ molecule pierces BP through the geometrical center of the heptagon (red dot shown in the leftmost inset, Figure \ref{fig:fig1}a) is  slightly smaller than 6.0 eV and accessible via light-induced excitations in the near ultraviolet range.  In computing these energy barriers with spin-polarized calculations in which the total spin is left unrestricted, the spin of O$_2$ transitions from a triplet state at $|z|\gtrsim$ 2 \AA{} of the center of mass of the BP monolayer, to a singlet state for $|z|\lesssim$ 2 \AA{}. This change of spin configuration of the optimal structures is in agreement with previous reports \cite{oxygenplanar1}.

The reduction of the energy barrier at this intrinsic defect is naturally related to the larger metric invariants (Tr(g) and Det(g) seen in the inset of Figure 1a) that lower the electronic repulsion significantly. Crucially, this is only one of many possible structural defects on BP, as this two-dimensional material is known to be polymorphic.\cite{Liang,edges2,Tomanek1,Tomanek2,Tomanek3,Wu,Liu}

The effect of curvature on $E_a$ is studied in a conical structure that contains about 500 atoms and acquires its largest curvature and compressive strain at its apex, as indicated by the white tones on Tr(g) and Det(g) and the large values of $H$ and $K$.\cite{us5} A dimer piercing the apex (upper-left inset on Figure \ref{fig:fig1}b) encounters a maximum value of $E_a$  close to 18 eV, a value further away from the reach of optically-induced chemistry. The increase of the barrier arises from the structural compression that accompanies the creation of curvature at the apex.\cite{us5} One can tell ``up'' from ``down'' on a conical structure, and this distinction makes $E_a(z)$ asymmetric (solid, asymmetric black line on Figure \ref{fig:fig1}b).

The conical structure employed in Figure \ref{fig:fig1}b has a dislocation/disclination axis.\cite{us5} The characteristic structural reconstruction at the edges of BP\cite{Liang,edges2} is a manifestation of the polymorphism of this material,\cite{Tomanek1,Tomanek2,Tomanek3,Wu} and we found a metastable local structure along the disclination line of the conical structure with two atoms pulled away from a common bond during the structural optimization. We envision that similar defects may originate on BP during growth. This metastable structure does not localize electronic states within the semiconducting gap either,\cite{Liang} a fact that will later be discussed in greater detail. The barrier $E_a$ is very low at this defect, largely overlapping with the energy of photons in the visible range, as depicted by the yellow rectangle on Figure \ref{fig:fig1}b. An oxygen dimer attaching to BP at this intrinsic defect requires an activation energy smaller than 4.0 eV, which is available through an electronic photo-excitation induced by violet light. The accessible energy barrier for photo-oxidation at this second defect provides conclusive validation of the hypothesis that the photo-oxidation and degradation of BP originates at intrinsic defects.

\subsection{Chemisorption and dissociation of oxygen dimers}\label{sec:sec2}

\begin{figure}[tb]
\includegraphics[width=1.0\textwidth]{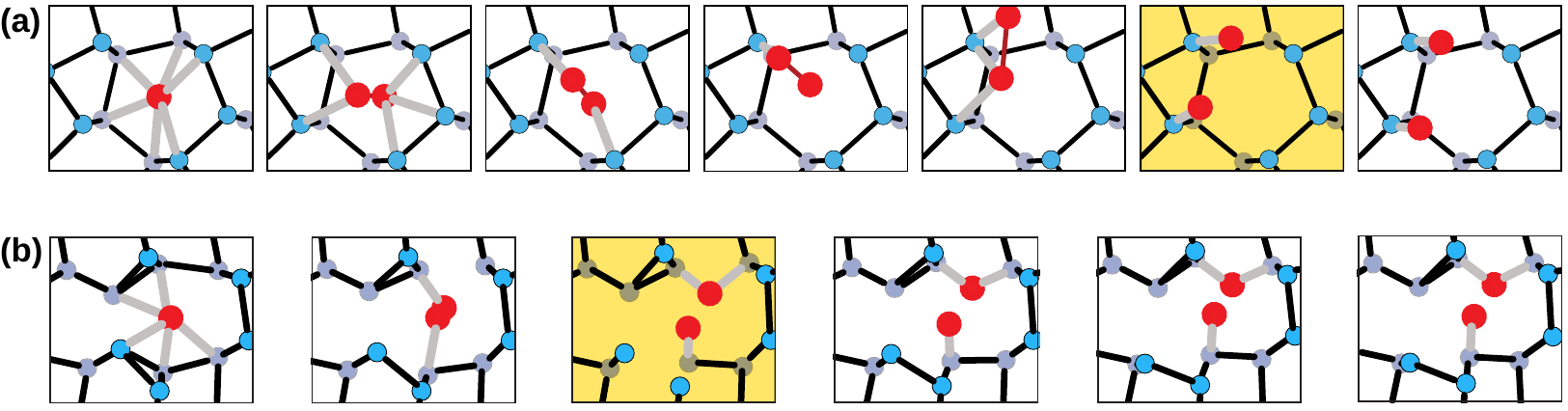}
\caption{Unveiling the dissociation of O$_2$ at intrinsic defects. The first frames to the left show the initial vertical placement of O$_2$ within the intrinsic defects. Oxygen dissociation at these intrinsic defects occurs at the frames highlighted in yellow. Frames at the far right show the optimized structures, at which distances among O atoms are 2.96 and 3.08 \AA{} for at structure (a) and (b), respectively. For reference, the equilibrium distance among O atoms in O$_2$ is 1.25 \AA. Full movies of the dissociation process are available as Supporting Information.}
\label{fig:figreel}
\end{figure}

Once an O$_2$ molecule is placed within an intrinsic defect aided by a suitable optical excitation, it will initially be expelled from BP with a force that has a normal component proportional to the slope of the curves in Figure \ref{fig:fig1}: $F_z=-\partial{E_a}/\partial z$,  and forces felt by individual oxygen atoms lead to the dissociation of O$_2$ near BP at intrinsic defects. We demonstrate this dissociation mechanism next.

We registered the atomistic dynamics of O$_2$ placed  initially at rest into the two intrinsic defects displaying the smallest energy barriers in Figure \ref{fig:fig1}   and assuming that the atomistic reconfiguration is adiabatic. (Under non-adiabatic conditions, the dimer may collect sufficient kinetic energy to cross onto a subsequent monolayer.) Full movies showing this process (movies S1 and S2) can be found as Supporting Information. The fundamental finding is that the distance among O atoms increases from 1.25 up to 2.96 \AA{} at the end of the structural optimization in Figure 2a, and up to 3.08 \AA{} at the end of the optimization depicted in Figure 2b: O$_2$ dissociates at intrinsic defects. Additional details follow.

As indicated previously, the dimer was placed vertically at the onset of the optimization, but it is expelled from the defect as the O atoms end up dissociated and lining up horizontally. Distances from O atoms to the closest P atoms are equal to 1.53 \AA{} and the P-P-O angles subtended to the three closest P atoms are equal to 111$^{\circ}$, 116$^{\circ}$, and 124$^{\circ}$.

The oxidation mediated by the second intrinsic defect provides further evidence for the fact that defects help dissociate O$_2$ due to a non-symmetric force that pulls O atoms apart. The first dissociated O atom in Figure \ref{fig:figreel}b binds to a single P atom with a bond distance of 1.56 \AA{}, quite similar to the distance seen for the O-P single bonds in Figure \ref{fig:figreel}a. The second O atom binds to two P atoms with distances of 1.73 \AA. We address the energetics as BP oxidizes next.

\subsection{The energetics of oxidation}

\begin{figure}[tb]
\includegraphics[width=1.0\textwidth]{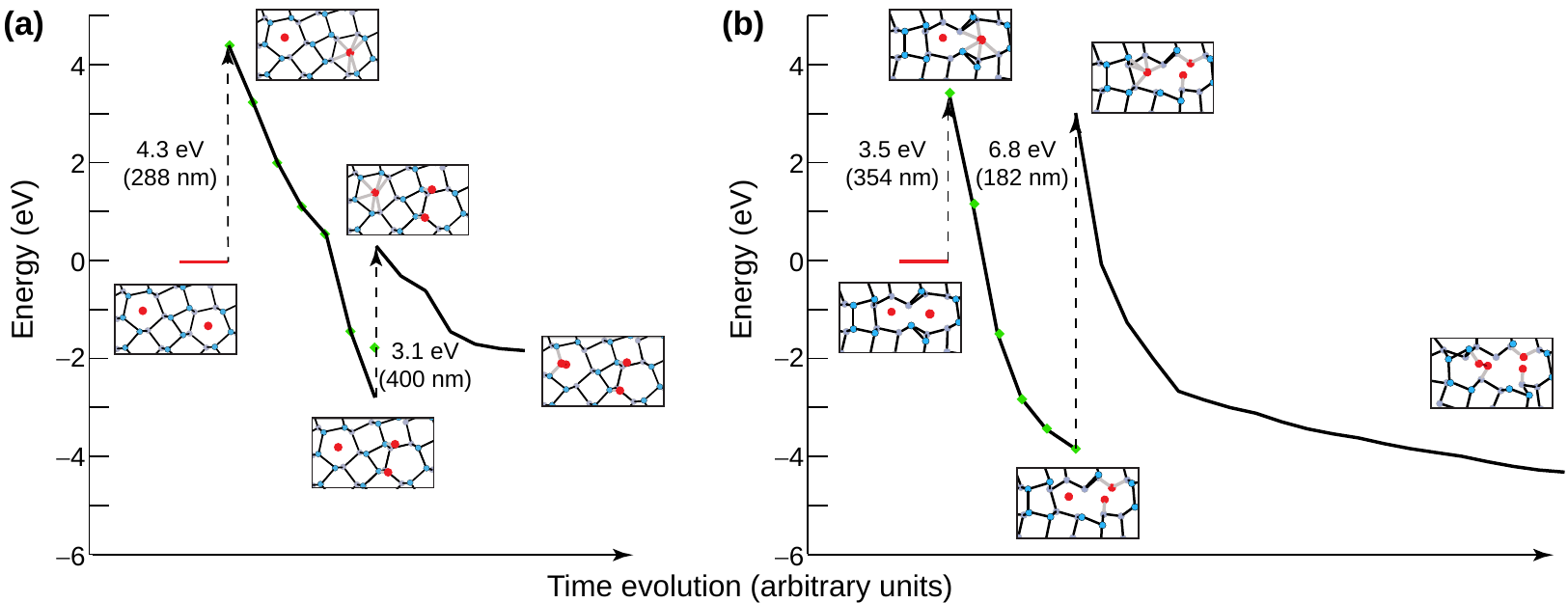}
\caption{Energetics at the onset of oxidation. The structures gain between  3.1 and 6.8 eV upon absorption of individual oxygen dimers, and release from 1.9 to about 7.0 eV upon oxidation, indicating how favorable oxidation is once the initial absorption barriers are accessed with the aid of optically-induced excitations. Green diamonds indicate the energies at the snapshots displayed in Figure \ref{fig:figreel}.}
\label{fig:figenergetics}
\end{figure}

We now study the successive oxidation of two dimers at nearby structural defects. The two dimers are placed far above BP prior to oxidation in order to set a suitable reference energy. As the first dimer is placed into the pentagon defect in Figure \ref{fig:figenergetics}a, it binds to BP and the system gains 4.3 eV. The system is allowed to relieve forces (see Methods) and it releases an energy of about 7 eV in the process, making for a highly exothermic reaction.\cite{oxygenplanar2,oxygenplanar4} Such a large energy release could unleash a continued  oxidation process --even in the absence of external illumination-- if not promptly dissipated.

Continuing the discussion of Figure \ref{fig:figenergetics}a, a second dimer is placed into BP at a second heptagonal defect, gaining  3.1 eV as it binds to BP. It is energetically unfavorable for oxygen dimer to oxidize consecutive heptagonal defects on this dislocation line, as the relaxed structure containing two dimers has an energy larger than the structure in which the second dimer is yet to be absorbed. Oxygens on this second dimer did not dissociate, but the distance between the two P atoms increases from 2.3 \AA{} to 3.2 \AA{} as a common P-P bond is broken; these two P atoms bind to a single O atom with bond distances of 1.87 \AA{} and a P-O-P angle of 117$^{\circ}$. Oxygens on the second dimer remain bounded at a larger separation of 1.48 \AA{}. There are no significant rearrangements of the first two O atoms upon absorption of the second dimer.

The process is repeated for two O$_2$ molecules that oxidize the structural defect that was shown in Figures \ref{fig:fig1}b and \ref{fig:figreel}b. As seen on Figure \ref{fig:figenergetics}b, the energy cost upon absorption of the first dimer is  3.5 eV, which could be accessible if the system adsorbs light with a wavelength of 354 nm. An energy release of 7.0 eV is recorded again as the structural forces are relieved. The second dimer seen in Figure \ref{fig:figenergetics}b is unable to dissociate and has a distance of 1.49 \AA{} among O atoms, and of 1.80 \AA{} among O-P bonds. The structure gains 6.8 eV when the second dimer is absorbed, which may be enabled by an optical excitation in the ultraviolet range, and releases about 7.0 eV through the atomistic optimization process to make for a slightly exothermic reaction (see Methods).  The results from Figure 3 indicate that larger energy gains can be attained when the second dimer is not in close proximity to the first absorbed O$_2$ molecule.

The substantial energy gains upon chemisorption recorded in Figure \ref{fig:figenergetics} imply that the oxidized structures are more stable than BP containing intrinsic defects, providing a clear picture of how reactive the material can be at these defects. Even though the defects studied in Figure \ref{fig:figenergetics}a and \ref{fig:figenergetics}b are quite different, they exhibit a similar trend in the oxidation barriers and in the energy release upon oxidation, an encouraging finding that enables a general understanding of the oxidation of BP in terms of energy barriers alone, regardless of the specific atomistic arrangements found at individual intrinsic defects.

\subsection{Geometrical and electronic characterization of the oxidized structures}

\begin{figure}[tb]
\includegraphics[width=1.0\textwidth]{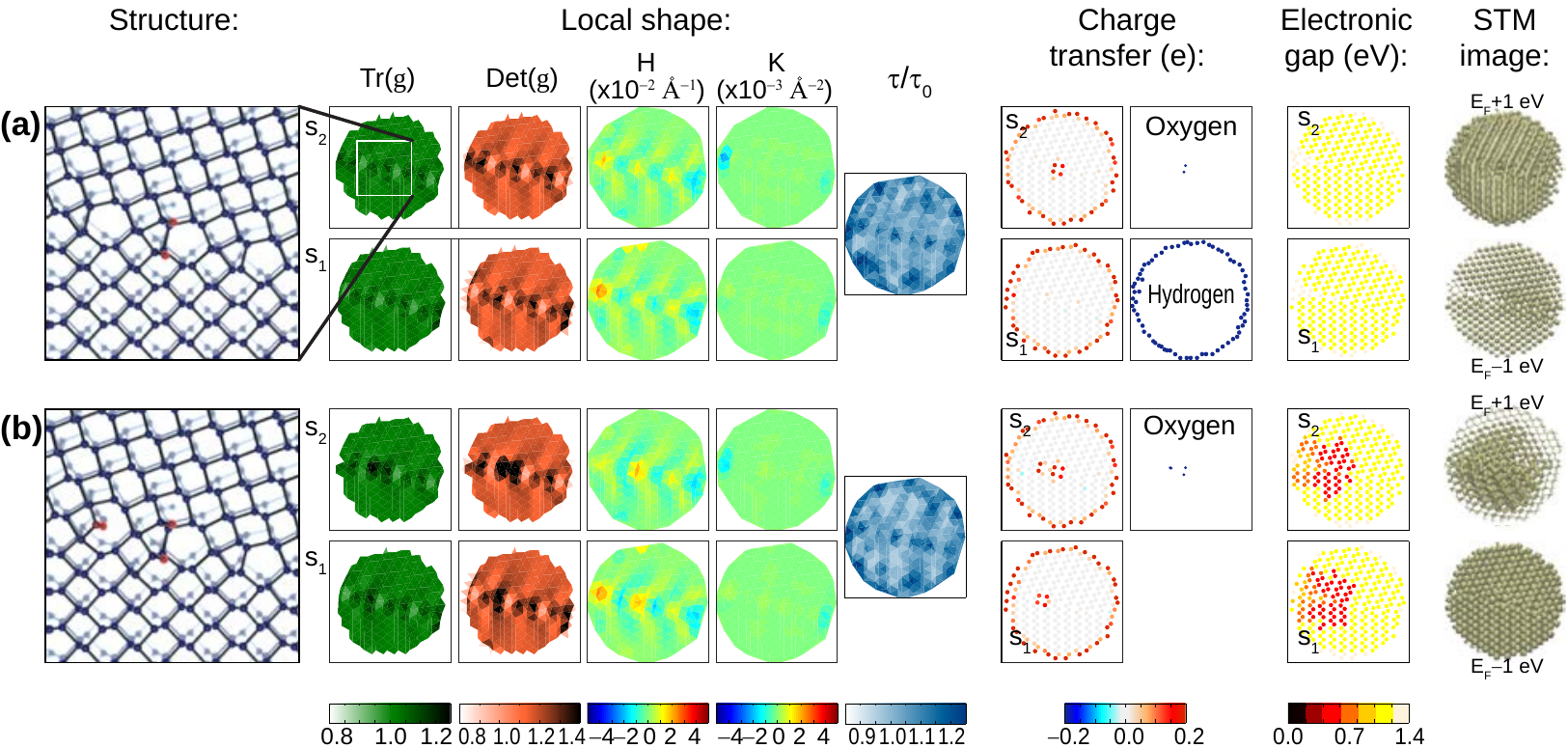}
\caption{Local shape and electronic properties of the planar BP structure as it absorbs (a) the first and (b) the second oxygen dimer. The ``structure'' subplots represent a segment of the material larger than the one seen in Figures \ref{fig:fig1}a, \ref{fig:figreel}a and \ref{fig:figenergetics}a. BP has a periodically buckled structure, and the geometry is displayed at sublayers $s_1$ and $s_2$ along with the local thickness $\tau/\tau_0$. Oxygen dimers (and hydrogen atoms at the finite boundaries of the BP monolayer) gain electrons from BP and the electronic gap is reduced as the second dimer is absorbed. Simulated STM images are shown too.}
\label{fig:fig4}
\end{figure}

\begin{figure}[tb]
\includegraphics[width=1.0\textwidth]{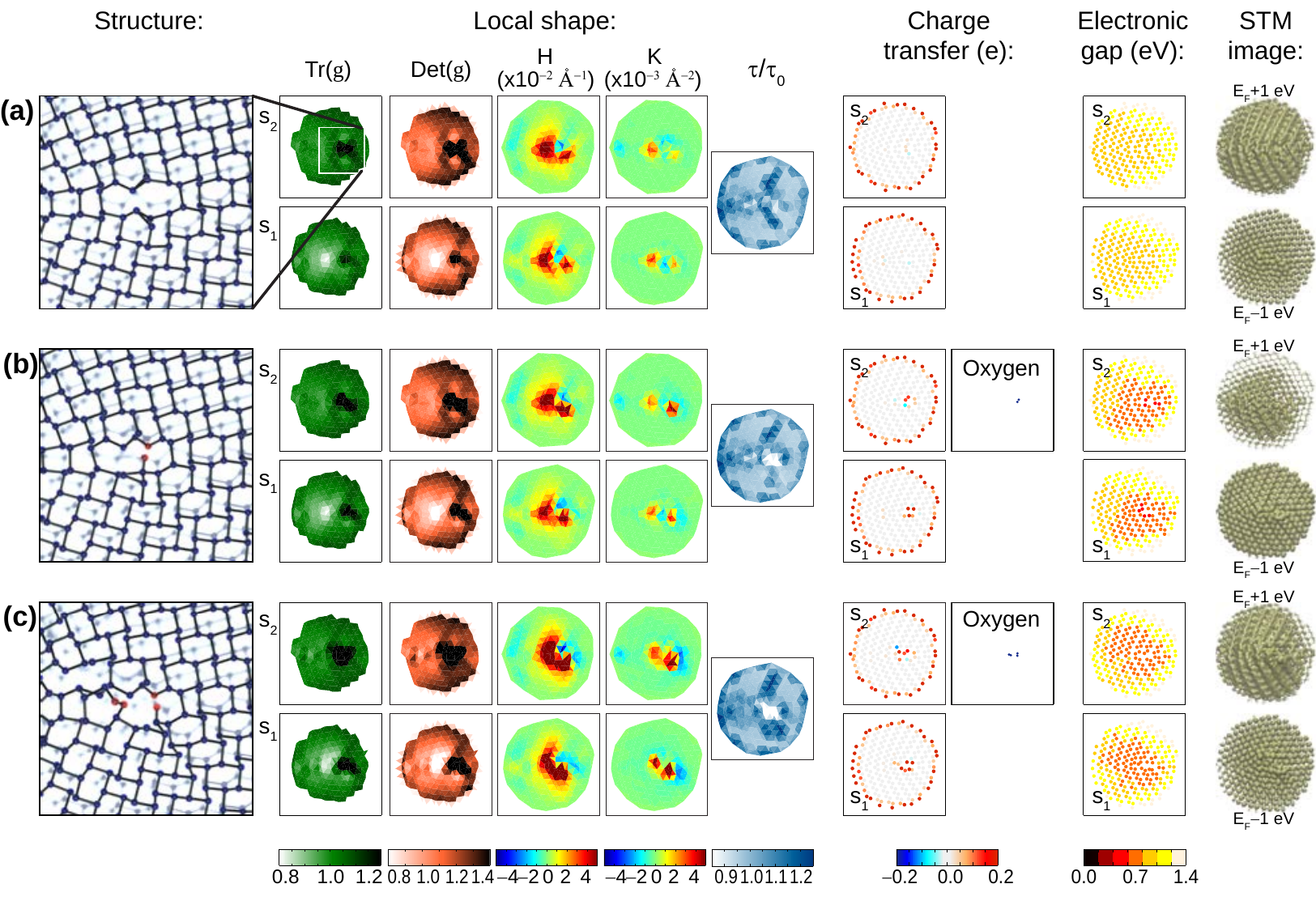}
\caption{
Conical structure (a) without oxygen dimers; (b) with one dimer; and (c) with two dimers. The shape information now indicates a clear tendency towards amorphization as BP oxidizes: BP sees fluctuations on its metric and curvature, and it bulges too as seen on the $\tau/\tau_0$ subplots. Oxygen atoms gain electrons from BP, and the electronic gap is reduced as these two dimers are absorbed. Simulated STM images are also shown.}
\label{fig:fig5}
\end{figure}

We conclude this work with an analysis of the global shape and the electronic properties of the oxidized structures that is provided in Figures \ref{fig:fig4} and \ref{fig:fig5}. A line defect\cite{Yakobson} does not produce significant curvature on BP ($H\simeq 0$ and $K\simeq 0$ in Figure \ref{fig:fig4}), but it creates a periodic compression/elongation of interatomic distances that is captured by the white/black tones seen on Tr(g) and Det(g) in Figure \ref{fig:fig4}.\cite{us5} The squares on the individual {\em local shape} subplots on this Figure have sides that are 50 \AA{} long and provide a global view of the shape of the finite systems we work with.

The chemisorption of a single oxygen dimer does not change the shape of this planar structure\cite{Yakobson} significantly; as indicated in previous paragraphs, BP accommodated the dimer by bringing it out of the planar structure and thus maintaining its original shape without any significant alteration (see {\em local shape} subplots in Figure \ref{fig:fig4}a). The oxygen atoms receive a large amount of electronic charge from phosphorus atoms (see {\em charge transfer} subplot in Figure \ref{fig:fig4}a). Additionally, the hydrogen atoms that passivate the edges of the final structure receive electronic charge from phosphorus. However, a similar charge transfer from hydrogen atoms is seen throughout all systems studied, and for that reason these plots are omitted from now on.

The next question is whether localized states that decrease the magnitude of the electronic gap are created at early stages of oxidation. This question is answered by determining the first electronic wavefunctions below and above the Fermi energy ($E_F$) having a non-zero density at a given individual atom. A single chemisorbed O$_2$ does not decrease the magnitude of the electronic gap (see {\em electronic gap} subplot), which remains close to about 1.1 eV in Figure \ref{fig:fig4}a due to finite-size effects\cite{us5}.

Many two-dimensional materials localize electronic states at intrinsic defects. In fact, it is the presence of these localized electronic states what permits the facile identification of structural defects.\cite{Kotakoski}  But many structural reconstructions of BP \cite{Yakobson,Liang} lack localized electronic states, making the identification of defects under TEM and optical probes a more difficult task. The simulated STM images provided at the far right of Figure \ref{fig:fig4}a are three-dimensional isosurface images produced from the density of electronic states having energies up to +1 (down to $-$1) eV from $E_F$.\cite{deg7,NanoLetters} The dislocation line is apparent on the STM at +1 eV. The patterns seen on the STM image at energies below $E_F$ are also seen on the {\em electronic gap} subplot. All remaining Figures in this manuscript were arranged following a layout identical to that of Figure \ref{fig:fig4}a.

The electronic properties of the planar BP sample after the chemisorption of a second dimer are shown in Figure \ref{fig:fig4}b. The increase in distances among P atoms once the second dimer was absorbed is captured by a new visible dark pattern on the metric invariants in the upper sublayer $s_2$ on Figure \ref{fig:fig4}b where the second dimer was absorbed. Curvatures ($H$ and $K$ on Figure \ref{fig:fig4}b) continue to be negligible and the ratio $\tau/\tau_0$ remains close to unity, indicating no significant bulging either. Importantly, the overall shape observed in Figure \ref{fig:fig4} tells us that the absorption of oxygen does not induce a significant curvature at the onset of the oxidation process. Oxygen atoms continue to gain electronic charge as the second dimer is added. A significant difference, though, concerns the magnitude of the gap that is reduced this time, as highlighted by the red undertones in the {\em electronic gap} subplot in Figure \ref{fig:fig4}b. This localized charge is also evident on the simulated iso-surface plots.

If the structure has some curvature originally, the oxidation process tends to increase it. To demonstrate this statement, the conical structure has a modest protrusion of 4 \AA{} and a slant height of about 25 \AA{} (for the sake of comparison, a graphene cone will have a much larger height equal to 14 \AA{} for a slant height of 25 \AA{}). This variation in height is well within the range of heights measured for the exposed layer on recently exfoliated black phosphorus stacks.

The initial shape of the conical structure studied in Figures \ref{fig:fig1}b, \ref{fig:figreel}b, and \ref{fig:figenergetics}b, can be seen on Figure \ref{fig:fig5}a and has the following salient features: the lower (upper) sublayer $s_1$ ($s_2$) is compressed (elongated) near the apex, as indicated by the white (black) features on Tr(g) and Det(g) around the structure's geometrical center. The self-passivated defect is shown in black on Tr(g) and Det(g) to the right of the geometrical center, indicating increased metric invariants there. This defect did not pin electronic charge within the electronic semiconducting gap,\cite{Yakobson,Liang} as can be seen on the {\em electronic gap} subplot in Figure \ref{fig:fig5}a. As a result, the band gap remains of the order of 0.89 eV down from 1.1 eV on the planar structure in an effect due to curvature.\cite{us5} The lack of localized states at the defect\cite{Yakobson,Liang,edges2} may mask its identification by optical probes.

The absorption of O$_2$ in a planar structure did not increase the curvature significantly, but there is a clear increase in the local magnitude of the gaussian curvature $K$ in the conical structure, Figure \ref{fig:fig5}b, as soon as the dimer is absorbed: oxidation increases the curvature of an already non-ideal structure. In terms of experiments, this means that a structure with a large height distribution displays local changes of curvature as well, and the bulges seen experimentally after degradation are consistent with an increase of curvature upon oxidation. The gradual increase of curvature is reaffirmed in Figure \ref{fig:fig5}c. The electronic gap of this structure reduces, somewhat dramatically, to 0.65 eV --signifying a 27\% decrease with respect to the gap of the structure seen in Figure \ref{fig:fig5}a. An increase in amorphization\cite{oxygenplanar6} is evident when the second dimer is part of the BP structure (Figure \ref{fig:fig5}c), and the electronic properties do not seem to be largely modified from what it was discussed in Figure \ref{fig:fig5}b.

\subsection{Conclusion}
The search for mechanisms to explain the oxidation of black phosphorus remains one of the most exciting and challenging problems in the context of this layered material where chemistry plays a fundamental role. We provided a viable mechanism for the photo-oxidation of BP at intrinsic defects. Intrinsic defects lower the chemisorption barrier of ideal black phosphorus and make the oxidation barrier accessible through the absorption of photons within the visible and ultraviolet range, thus enabling the photo-induced oxidation of BP and the dissociation of oxygen dimers. We studied energetics of chemisorption and found the oxidation to be highly exothermic. In addition, the local geometry, electronic properties, and simulated STM images of these structures were discussed as well. Further avenues of study concern the degradation of subsequent layers following similar mechanisms. This study provides novel insights into the degradation dynamics of this material, and gives rise to a plausible mechanism for the oxidation of black phosphorus when exposed to light.

\subsection{Methods}
Structural optimizations were carried out with the {\em SIESTA} DFT code\cite{Car,SIESTA1,SIESTA2} with spin-polarized density functional theory and the PBE exchange-correlation potential, following a conjugate-gradient method until all atomic force components were smaller than 0.04 eV/\AA. All P and O atoms were allowed to relax their forces during the optimization process. Movies displaying the chemisorption process were created from individual coordinate snapshots as the conjugate-gradient optimization went on, following an in-house scripting procedure. The analysis of the discrete geometrical conformations is adapted from Ref.~\cite{conf} (see Ref.~\cite{DDG} too). The ``local gaps'' are obtained by projecting the first state having a non-zero electronic density at any given atom below and above the Fermi level, and are indicative of charge localization and level pinning within the electronic gap. STM images were created from individual wavefunctions whose density is squared and added up (down) 1 eV from $E_F$.\cite{NanoLetters,deg7}

To define the discrete geometry employed on Figures 1, 4 and 5, we consider three {\em directed} edges $\mathbf{e}_1$, $\mathbf{e}_2$ and $\mathbf{e}_3$ such that $\mathbf{e}_1+\mathbf{e}_2+\mathbf{e}_3=\mathbf{0}$, and define $Q_l^I\equiv\mathbf{e}_l\cdot\mathbf{e}_l$ ($l=1,2,3$), representing the square of the smallest {\em finite}  distance among atoms on the 2-D lattice \cite{conf}.

\begin{figure}
\centerline{\includegraphics[width=0.5\textwidth]{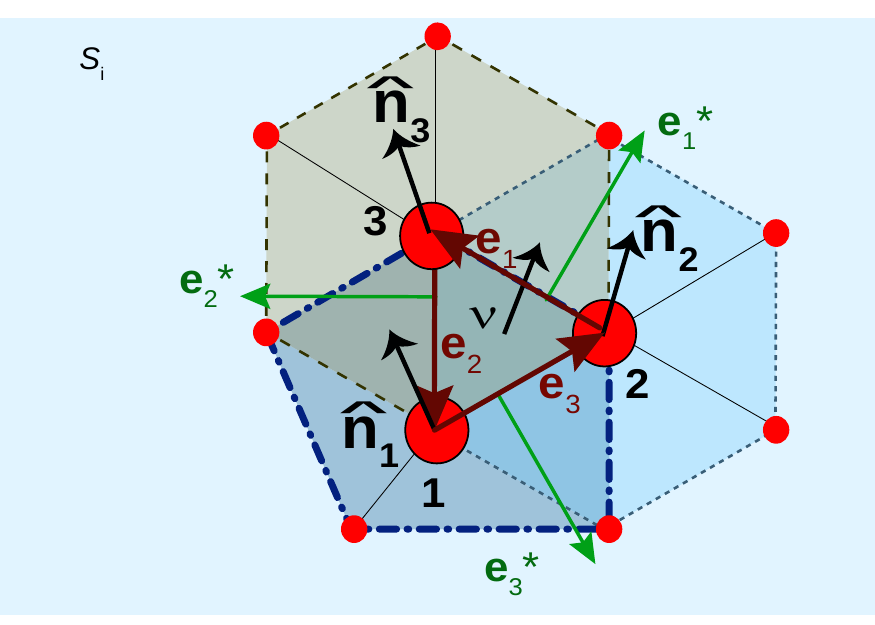}}\label{fig:F6}
\caption{Discrete tensors based on triangulations are expressed in terms of averaged normals $\hat{\mathbf{n}}_j$, edges $\mathbf{e}_j$, the normal of the triangle $\boldsymbol{\nu}$, and dual edges $\mathbf{e}_j^*\equiv\boldsymbol{\nu}\times \mathbf{e}_j$ ($j=1,2,3$).}
\end{figure}

We consider the change in orientation among normals $\hat{\mathbf{n}}_j$ and $\hat{\mathbf{n}}_k$ as well, and project such variation onto their common edge $\mathbf{e}_l$: This is is, one defines $Q_l^{II}\equiv (\hat{\mathbf{n}}_k-\hat{\mathbf{n}}_j)\cdot \mathbf{e}_l$ (see Figure~6; $j,k,l$ are permutations of integers 1, 2, and 3). In previous Equation, $\hat{\mathbf{n}}_l$ is the average over individual normals at triangulated area elements within the polygon surrounding atom $l$ and highlighted by dashed lines on Figure 6. The dual edge is defined by $\mathbf{e}_l^*\equiv \boldsymbol{\nu}\times \mathbf{e}_l$, with $\boldsymbol{\nu}$ the normal to the triangle formed by atoms $\mathbf{1}$, $\mathbf{2}$ and $\mathbf{3}$ and $A_T$ is the triangle area ($-\mathbf{e}_1\times \mathbf{e}_2=2A_T\boldsymbol{\nu}$) \cite{conf}.

This way, the discrete metric tensor takes the following form \cite{us5,conf}:
\begin{equation}\label{eq:eq8}
g=-\frac{1}{8A_0^2}\sum_{(j,k,l)}(Q_j^I - Q_k^I- Q_l^I)\mathbf{e}_j^*\otimes\mathbf{e}_j^*,
\end{equation}
with $A_0$ the area of the triangulated area element at the reference (non-deformed and defect-free) plane.

The discrete curvature tensor is:
\begin{equation}\label{eq:eq7}
k=-\frac{1}{8A_T^2}\sum_{(j,k,l)}(Q_j^{II}-Q_k^{II}-Q_l^{II})\mathbf{e}_j^*\otimes\mathbf{e}_j^*.
\end{equation}
The parenthesis $(j,k,l)$ indicates a sum of three terms, as follows: ($j=1$, $k=2$, $l=3$), (2, 3, 1), and (3, 1, 2). Eqns.~\eqref{eq:eq8} and \eqref{eq:eq7}  become 3$\times$3 matrices with explicit values for $Q_l^I$, $Q_l^{II}$ and $\mathbf{e}_j^*$ from atomic positions. For instance, the discrete curvature tensor has eigenvalues \{0, $k_1$, $k_2$\} at each triangulated area element, yielding $H=(k_1+k_2)/2$ and $K=k_1k_2$. The geometrical invariants reported at point $\mathbf{j}$ are averages over their values at individual triangles sharing this vertex.

The thickness $\tau$ is the distance among an atom in the lower sublayer $S_1$ and the centroid of three nearest atoms belonging to the upper sublayer $S_2$.

\section{Associated content}
\subsection{Supporting information}
Movies showing the dissociation of oxygen dimers at BP are available free of charge n the ACS Publications website.

\subsection{Author contributions}
K.U., P.R., M.F.B. and S.B.L. computed the absorption barriers and performed Car-Parinello molecular dynamics. A.A.P.S. and E.O.H. contributed the geometrical analysis. P.R. and S.B.L. provided the study of electronic properties and the simulated STM images. S.B.L. wrote the paper with input from the other authors.

\subsection{Notes}
The authors declare no competing financial interest.

\section{Acknowledgements}
Calculations were performed at the High Performance Computing Centers at Arkansas and Oklahoma State University (NSF, Grant OCI-1126330), and on TACC's {\em Stampede} (NSF-XSEDE ACI-1053575 and TG-PHY090002).


\begin{mcitethebibliography}{71}
\providecommand*\natexlab[1]{#1}
\providecommand*\mciteSetBstSublistMode[1]{}
\providecommand*\mciteSetBstMaxWidthForm[2]{}
\providecommand*\mciteBstWouldAddEndPuncttrue
  {\def\EndOfBibitem{\unskip.}}
\providecommand*\mciteBstWouldAddEndPunctfalse
  {\let\EndOfBibitem\relax}
\providecommand*\mciteSetBstMidEndSepPunct[3]{}
\providecommand*\mciteSetBstSublistLabelBeginEnd[3]{}
\providecommand*\EndOfBibitem{}
\mciteSetBstSublistMode{f}
\mciteSetBstMaxWidthForm{subitem}{(\alph{mcitesubitemcount})}
\mciteSetBstSublistLabelBeginEnd
  {\mcitemaxwidthsubitemform\space}
  {\relax}
  {\relax}

\bibitem[Novoselov et~al.(2005)Novoselov, Jiang, Schedin, Booth, Khotkevich,
  Morozov, and Geim]{2D1}
Novoselov,~K.~S.; Jiang,~D.; Schedin,~F.; Booth,~T.~J.; Khotkevich,~V.~V.;
  Morozov,~S.~V.; Geim,~A.~K. Two-dimensional atomic crystals. \emph{Proc. Natl. Acad. Sci. (USA)}
  \textbf{2005}, \emph{102}, 10451–10453\relax
\mciteBstWouldAddEndPuncttrue
\mciteSetBstMidEndSepPunct{\mcitedefaultmidpunct}
{\mcitedefaultendpunct}{\mcitedefaultseppunct}\relax
\EndOfBibitem
\bibitem[Butler et~al.(2013)Butler, Hollen, Cao, Cui, Gupta, Guti{\'e}rrez,
  Heinz, Hong, Huang, Ismach, Johnston-Halperin, Kuno, Plashnitsa, Robinson,
  Ruoff, Salahuddin, Shan, Shi, Spencer, Terrones, Windl, and Goldberger]{2D2}
Butler,~S.~Z. et~al.  Progress, challenges, and opportunities in two-dimensional materials beyond graphene. \emph{ACS Nano} \textbf{2013}, \emph{7}, 2898–2926\relax
\mciteBstWouldAddEndPuncttrue
\mciteSetBstMidEndSepPunct{\mcitedefaultmidpunct}
{\mcitedefaultendpunct}{\mcitedefaultseppunct}\relax
\EndOfBibitem
\bibitem[{Castro~Neto} et~al.(2009){Castro~Neto}, Guinea, Peres, Novoselov, and
  Geim]{gr1}
{Castro~Neto},~A.~H.; Guinea,~F.; Peres,~N.~M.~R.; Novoselov,~K.~S.;
  Geim,~A.~K. The electronic properties of graphene. \emph{Rev. Mod. Phys.} \textbf{2009}, \emph{81}, 109-162\relax
\mciteBstWouldAddEndPuncttrue
\mciteSetBstMidEndSepPunct{\mcitedefaultmidpunct}
{\mcitedefaultendpunct}{\mcitedefaultseppunct}\relax
\EndOfBibitem
\bibitem[Katsnelson(2012)]{gr2}
Katsnelson,~M.~I. \emph{Graphene: Carbon in two dimensions}, 1st ed.;
  Cambdridge U. Press: Cambridge, 2012\relax
\mciteBstWouldAddEndPuncttrue
\mciteSetBstMidEndSepPunct{\mcitedefaultmidpunct}
{\mcitedefaultendpunct}{\mcitedefaultseppunct}\relax
\EndOfBibitem
\bibitem[Watanabe et~al.(2004)Watanabe, Taniguchi, and Kanda]{hbn1}
Watanabe,~K.; Taniguchi,~T.; Kanda,~H. Direct-bandgap properties and evidence for ultraviolet lasing of hexagonal boron nitride single crystal. \emph{Nature Mater.} \textbf{2004},
  \emph{3}, 404-409\relax
\mciteBstWouldAddEndPuncttrue
\mciteSetBstMidEndSepPunct{\mcitedefaultmidpunct}
{\mcitedefaultendpunct}{\mcitedefaultseppunct}\relax
\EndOfBibitem
\bibitem[Jin et~al.(2009)Jin, Lin, Suenaga, and Iijima]{hbn2}
Jin,~C.; Lin,~F.; Suenaga,~K.; Iijima,~S. Fabrication of a freestanding boron nitride single layer and its defect assignments. \emph{Phys. Rev. Lett.}
  \textbf{2009}, \emph{102}, 195505\relax
\mciteBstWouldAddEndPuncttrue
\mciteSetBstMidEndSepPunct{\mcitedefaultmidpunct}
{\mcitedefaultendpunct}{\mcitedefaultseppunct}\relax
\EndOfBibitem
\bibitem[Wang et~al.(2012)Wang, Kalantar-Zadeh, Kis, Coleman, and Strano]{TMD1}
Wang,~Q.~H.; Kalantar-Zadeh,~K.; Kis,~A.; Coleman,~J.~N.; Strano,~M.~S. Electronics and optoelectronics of two-dimensional transition metal dichalcogenides. \emph{Nature Nanotechnol.} \textbf{2012}, \emph{7}, 699-712\relax
\mciteBstWouldAddEndPuncttrue
\mciteSetBstMidEndSepPunct{\mcitedefaultmidpunct}
{\mcitedefaultendpunct}{\mcitedefaultseppunct}\relax
\EndOfBibitem
\bibitem[Chowalla et~al.(2013)Chowalla, Shin, Eda, Li, Loh, and Zhang]{TMD2}
Chowalla,~M.; Shin,~H.~S.; Eda,~G.; Li,~L.~J.; Loh,~K.~P.; Zhang,~H. The chemistry of two-dimensional layered transition metal dichalcogenide nanosheets. \emph{Nature Chem.} \textbf{2013}, \emph{5}, 263–275\relax
\mciteBstWouldAddEndPuncttrue
\mciteSetBstMidEndSepPunct{\mcitedefaultmidpunct}
{\mcitedefaultendpunct}{\mcitedefaultseppunct}\relax
\EndOfBibitem
\bibitem[Najmaei et~al.(2013)Najmaei, Liu, Zhou, Zou, Shi, Lei, Yakobson,
  Idrobo, Ajayan, and Lou]{TMD3}
Najmaei,~S.; Liu,~Z.; Zhou,~W.; Zou,~X.; Shi,~G.; Lei,~S.; Yakobson,~B.~I.;
  Idrobo,~J.; Ajayan,~P.~M.; Lou,~J. Vapour phase growth and grain boundary structure of molybdenum disulphide atomic layers. \emph{Nature Mater.} \textbf{2013},
  \emph{12}, 754–759\relax
\mciteBstWouldAddEndPuncttrue
\mciteSetBstMidEndSepPunct{\mcitedefaultmidpunct}
{\mcitedefaultendpunct}{\mcitedefaultseppunct}\relax
\EndOfBibitem
\bibitem[{van~der~Zande} et~al.(2013){van~der~Zande}, Huang, Chenet,
  Berkelbach, You, Lee, Heinz, Reichman, Muller, and Hone]{TMD4}
{van~der~Zande},~A.~M.; Huang,~P.~Y.; Chenet,~D.~A.; Berkelbach,~T.~C.;
  You,~Y.; Lee,~G.~H.; Heinz,~T.~F.; Reichman,~D.~R.; Muller,~D.~A.;
  Hone,~J.~C. Grains and grain boundaries in highly crystalline monolayer molybdenum disulphide. \emph{Nature Mater.} \textbf{2013}, \emph{12}, 554–561\relax
\mciteBstWouldAddEndPuncttrue
\mciteSetBstMidEndSepPunct{\mcitedefaultmidpunct}
{\mcitedefaultendpunct}{\mcitedefaultseppunct}\relax
\EndOfBibitem
\bibitem[Britnell et~al.(2013)Britnell, Riveiro, Eckmann, Jalil, Belle,
  Mishchenko, Kim, Gorbachev, Georgiou, Morozov, Grigorenko, Geim, Casriaghi,
  Neto, and Novoselov]{TMD5}
Britnell,~L.; Riveiro,~R.~M.; Eckmann,~A.; Jalil,~R.; Belle,~B.~D.;
  Mishchenko,~A.; Kim,~Y.~J.; Gorbachev,~R.~V.; Georgiou,~T.; Morozov,~S.~V.;
  Grigorenko,~A.~N.; Geim,~A.~K.; Casriaghi,~C.; Neto,~A.~H.~C.;
  Novoselov,~K.~S. Strong light-matter interactions in heterostructures of atomically thin films. \emph{Science} \textbf{2013}, \emph{340}, 1311-1314\relax
\mciteBstWouldAddEndPuncttrue
\mciteSetBstMidEndSepPunct{\mcitedefaultmidpunct}
{\mcitedefaultendpunct}{\mcitedefaultseppunct}\relax
\EndOfBibitem
\bibitem[Guti{\'e}rrez et~al.(2013)Guti{\'e}rrez, Perea-L{\'o}pez, El{\'\i}as,
  Berkdemir, Wang, Lv, L{\'o}pez-Ur{\'\i}as, Crespi, Terrones, and
  Terrones]{TMD6}
Guti{\'e}rrez,~H.~R.; Perea-L{\'o}pez,~N.; El{\'\i}as,~A.~L.; Berkdemir,~A.;
  Wang,~B.; Lv,~R.; L{\'o}pez-Ur{\'\i}as,~F.; Crespi,~V.~H.; Terrones,~H.;
  Terrones,~M. Extraordinary room-temperature photoluminescence in triangular WS$_2$ monolayers. \emph{Nano Lett.} \textbf{2013}, \emph{13}, 3447–3454\relax
\mciteBstWouldAddEndPuncttrue
\mciteSetBstMidEndSepPunct{\mcitedefaultmidpunct}
{\mcitedefaultendpunct}{\mcitedefaultseppunct}\relax
\EndOfBibitem
\bibitem[Mak et~al.()Mak, Lee, Hone, Shan, and Heinz]{TMD7}
Mak,~K.~F.; Lee,~C.; Hone,~J.; Shan,~J.; Heinz,~T.~F. Atomically thin MoS$_2$: a new direct-gap semiconductor. \emph{Phys. Rev. Lett.}
  \emph{105}, 136805\relax
\mciteBstWouldAddEndPuncttrue
\mciteSetBstMidEndSepPunct{\mcitedefaultmidpunct}
{\mcitedefaultendpunct}{\mcitedefaultseppunct}\relax
\EndOfBibitem
\bibitem[Li et~al.(2014)Li, Yu, Ye, Ge, Ou, Wu, Feng, Chen, and Zhang]{BP1}
Li,~L.; Yu,~Y.; Ye,~G.~J.; Ge,~Q.; Ou,~X.; Wu,~H.; Feng,~D.; Chen,~X.~H.;
  Zhang,~Y. Black phosphorus field-effect transistors. \emph{Nature Nanotechnol.} \textbf{2014}, \emph{9}, 372–377\relax
\mciteBstWouldAddEndPuncttrue
\mciteSetBstMidEndSepPunct{\mcitedefaultmidpunct}
{\mcitedefaultendpunct}{\mcitedefaultseppunct}\relax
\EndOfBibitem
\bibitem[Liu et~al.(2014)Liu, Neal, Zhu, Luo, Xu, Tom{\'a}nek, and Ye]{BP2}
Liu,~H.; Neal,~A.~T.; Zhu,~Z.; Luo,~Z.; Xu,~X.; Tom{\'a}nek,~D.; Ye,~P.~D. Phosphorene: an unexplored 2d semiconductor with a high hole mobility. \emph{ACS Nano} \textbf{2014}, \emph{8}, 4033–4041\relax
\mciteBstWouldAddEndPuncttrue
\mciteSetBstMidEndSepPunct{\mcitedefaultmidpunct}
{\mcitedefaultendpunct}{\mcitedefaultseppunct}\relax
\EndOfBibitem
\bibitem[Fei and Yang(2014)Fei, and Yang]{BP3}
Fei,~R.; Yang,~L. Strain-engineering the anisotropic electrical conductance of few-layer black phosphorus. \emph{Nano Lett.} \textbf{2014}, \emph{14}, 2884–2889\relax
\mciteBstWouldAddEndPuncttrue
\mciteSetBstMidEndSepPunct{\mcitedefaultmidpunct}
{\mcitedefaultendpunct}{\mcitedefaultseppunct}\relax
\EndOfBibitem
\bibitem[Ling et~al.(2015)Ling, Wang, Huang, Xia, and Dresselhaus]{BP4}
Ling,~X.; Wang,~H.; Huang,~S.; Xia,~F.; Dresselhaus,~M.~S. The renaissance of black phosphorus. \emph{Proc. Natl. Acad. Sci. (USA)} \textbf{2015}, \emph{112}, 4523–4530\relax
\mciteBstWouldAddEndPuncttrue
\mciteSetBstMidEndSepPunct{\mcitedefaultmidpunct}
{\mcitedefaultendpunct}{\mcitedefaultseppunct}\relax
\EndOfBibitem
\bibitem[Tran et~al.(2014)Tran, Soklaski, Liang, and Yang]{BP5}
Tran,~V.; Soklaski,~R.; Liang,~Y.; Yang,~L. Layer-controlled band gap and anisotropic excitons in few-layer black phosphorus. \emph{Physical Review B}
  \textbf{2014}, \emph{89}, 235319\relax
\mciteBstWouldAddEndPuncttrue
\mciteSetBstMidEndSepPunct{\mcitedefaultmidpunct}
{\mcitedefaultendpunct}{\mcitedefaultseppunct}\relax
\EndOfBibitem
\bibitem[Qiao et~al.(2014)Qiao, Kong, Hu, Yang, and Ji]{BP6}
Qiao,~J.; Kong,~X.; Hu,~Z.-X.; Yang,~F.; Ji,~W. High-mobility transport anisotropy and linear dichroism in few-layer black phosphorus. \emph{Nature Comm.}
  \textbf{2014}, \emph{5}, 5475\relax
\mciteBstWouldAddEndPuncttrue
\mciteSetBstMidEndSepPunct{\mcitedefaultmidpunct}
{\mcitedefaultendpunct}{\mcitedefaultseppunct}\relax
\EndOfBibitem
\bibitem[Castellanos-Gomez et~al.(2014)Castellanos-Gomez, Vicarelli, Prada,
  Island, Narasimha-Acharya, Blanter, Groenendijk, Buscema, Steele, Alvarez,
  Zandbergen, Palacios, and {vand~der~Zant}]{castellanos}
Castellanos-Gomez,~A.; Vicarelli,~L.; Prada,~E.; Island,~J.~O.;
  Narasimha-Acharya,~K.~L.; Blanter,~S.~I.; Groenendijk,~D.~J.; Buscema,~M.;
  Steele,~G.~A.; Alvarez,~J.~V.; Zandbergen,~H.~W.; Palacios,~J.~J.;
  {vand~der~Zant},~H.~S.~J. Isolation and characterization of few-layer black phosphorus. \emph{2D Mater.} \textbf{2014}, \emph{1},
  025001\relax
\mciteBstWouldAddEndPuncttrue
\mciteSetBstMidEndSepPunct{\mcitedefaultmidpunct}
{\mcitedefaultendpunct}{\mcitedefaultseppunct}\relax
\EndOfBibitem
\bibitem[Yau et~al.(1992)Yau, Moffat, Bard, Zhang, and Lerner]{deg0}
Yau,~S.-L.; Moffat,~T.~P.; Bard,~A.~J.; Zhang,~Z.; Lerner,~M.~M. STM of the (010) surface of orthorhombic phosphorus. \emph{Chem.
  Phys. Lett.} \textbf{1992}, \emph{198}, 383-388\relax
\mciteBstWouldAddEndPuncttrue
\mciteSetBstMidEndSepPunct{\mcitedefaultmidpunct}
{\mcitedefaultendpunct}{\mcitedefaultseppunct}\relax
\EndOfBibitem
\bibitem[Koenig et~al.(2014)Koenig, Doganov, Schmidt, {Castro~Neto}, and
  {\"O}zyilmaz]{deg1}
Koenig,~S.~P.; Doganov,~R.~A.; Schmidt,~H.; {Castro~Neto},~A.~H.;
  {\"O}zyilmaz,~B. Electric field effect in ultrathin black phosphorus. \emph{Appl. Phys. Lett.} \textbf{2014}, \emph{104},
  103106\relax
\mciteBstWouldAddEndPuncttrue
\mciteSetBstMidEndSepPunct{\mcitedefaultmidpunct}
{\mcitedefaultendpunct}{\mcitedefaultseppunct}\relax
\EndOfBibitem
\bibitem[Wood et~al.(2014)Wood, Wells, Jariwala, Chen, Cho, Sangwan, Liu,
  Lauhon, Marks, and Hersam]{deg2}
Wood,~J.~D.; Wells,~S.~A.; Jariwala,~D.; Chen,~K.-S.; Cho,~E.; Sangwan,~W.~K.;
  Liu,~X.; Lauhon,~L.~J.; Marks,~T.~J.; Hersam,~M.~C. Effective passivation of exfoliated black phosphorus transistors against ambient degradation. \emph{Nano Lett.}
  \textbf{2014}, \emph{14}, 6964–6970\relax
\mciteBstWouldAddEndPuncttrue
\mciteSetBstMidEndSepPunct{\mcitedefaultmidpunct}
{\mcitedefaultendpunct}{\mcitedefaultseppunct}\relax
\EndOfBibitem
\bibitem[Favron et~al.(2015)Favron, Gaufr{\`e}s, Fossard, Phaneuf-L'Heureux,
  Tang, L{\'e}vesque, Loiseau, Leonelli, Francoeur, and Martel]{Martel}
Favron,~A.; Gaufr{\`e}s,~E.; Fossard,~F.; Phaneuf-L'Heureux,~A.; Tang,~N.;
  L{\'e}vesque,~P.; Loiseau,~A.; Leonelli,~R.; Francoeur,~S.; Martel,~R. Photooxidation and quantum confinement effects in exfoliated black phosphorus. \emph{Nature Materials} \textbf{2015}, \emph{14}, 826–832\relax
\mciteBstWouldAddEndPuncttrue
\mciteSetBstMidEndSepPunct{\mcitedefaultmidpunct}
{\mcitedefaultendpunct}{\mcitedefaultseppunct}\relax
\EndOfBibitem
\bibitem[Kim et~al.(2015)Kim, Liu, Zhu, Kim, Wu, Tao, Dodabalapur, Lai, and
  Akinwande]{deg3}
Kim,~J.-S.; Liu,~Y.; Zhu,~W.; Kim,~S.; Wu,~D.; Tao,~L.; Dodabalapur,~A.;
  Lai,~K.; Akinwande,~D. Toward air-stable multilayer phosphorene thin-films and transistors. \emph{Sci. Rep.} \textbf{2015}, \emph{5}, 8989\relax
\mciteBstWouldAddEndPuncttrue
\mciteSetBstMidEndSepPunct{\mcitedefaultmidpunct}
{\mcitedefaultendpunct}{\mcitedefaultseppunct}\relax
\EndOfBibitem
\bibitem[Doganov et~al.(2015)Doganov, O'Farrell, Koenig, Yeo, Ziletti,
  Carvalho, Campbell, Coker, Watanabe, Taniguchi, {Castro~Neto}, and
  {\"Ozyilmaz}]{deg4}
Doganov,~R.~A.; O'Farrell,~E.~C.~T.; Koenig,~S.~P.; Yeo,~Y.; Ziletti,~A.;
  Carvalho,~A.; Campbell,~D.~K.; Coker,~D.~F.; Watanabe,~K.; Taniguchi,~T.;
  {Castro~Neto},~A.~H.; {\"Ozyilmaz},~B. Transport properties of pristine few-layer black phosphorus by van der Waals passivation in an inert atmosphere. \emph{Nature Comm.} \textbf{2015},
  \emph{6}, 6647\relax
\mciteBstWouldAddEndPuncttrue
\mciteSetBstMidEndSepPunct{\mcitedefaultmidpunct}
{\mcitedefaultendpunct}{\mcitedefaultseppunct}\relax
\EndOfBibitem
\bibitem[Doganov et~al.(2015)Doganov, Koenig, Yeo, Watanabe, Taniguchi, and
  {\"Ozyilmaz}]{deg5}
Doganov,~R.~A.; Koenig,~S.~P.; Yeo,~Y.; Watanabe,~K.; Taniguchi,~T.;
  {\"Ozyilmaz},~B. Transport properties of ultrathin black phosphorus on hexagonal boron nitride. \emph{Appl. Phys. Lett.} \textbf{2015}, \emph{106},
  083505\relax
\mciteBstWouldAddEndPuncttrue
\mciteSetBstMidEndSepPunct{\mcitedefaultmidpunct}
{\mcitedefaultendpunct}{\mcitedefaultseppunct}\relax
\EndOfBibitem
\bibitem[Gillgren et~al.(2015)Gillgren, Wickramaratne, Shi, Espiritu, Wang, Hu,
  Wei, Liu, Mao, Watanabe, Taniguchi, Bockrath, Barlas, Lake, and Lau]{deg6}
Gillgren,~N.; Wickramaratne,~D.; Shi,~Y.; Espiritu,~T.; Wang,~J.; Hu,~J.;
  Wei,~J.; Liu,~X.; Mao,~Z.; Watanabe,~K.; Taniguchi,~T.; Bockrath,~M.;
  Barlas,~Y.; Lake,~R.~K.; Lau,~C.~N. Gate tunable quantum oscillations in air-stable and high mobility few-layer phosphorene heterostructures. \emph{2D Mater.} \textbf{2015}, \emph{2},
  011001\relax
\mciteBstWouldAddEndPuncttrue
\mciteSetBstMidEndSepPunct{\mcitedefaultmidpunct}
{\mcitedefaultendpunct}{\mcitedefaultseppunct}\relax
\EndOfBibitem
\bibitem[Rivero et~al.(2015)Rivero, Horvath, Zhu, Guan, Tom{\'a}nek, and
  Barraza-Lopez]{deg7}
Rivero,~P.; Horvath,~C.~M.; Zhu,~Z.; Guan,~J.; Tom{\'a}nek,~D.;
  Barraza-Lopez,~S. Simulated scanning tunneling microscopy images of few-layer phosphorus capped by graphene and hexagonal boron nitride monolayers. \emph{Phys. Rev. B} \textbf{2015}, \emph{91}, 115413\relax
\mciteBstWouldAddEndPuncttrue
\mciteSetBstMidEndSepPunct{\mcitedefaultmidpunct}
{\mcitedefaultendpunct}{\mcitedefaultseppunct}\relax
\EndOfBibitem
\bibitem[Liu et~al.(2014)Liu, Xu, Zhang, Penev, and Yakobson]{Yakobson}
Liu,~Y.; Xu,~F.; Zhang,~Z.; Penev,~E.~S.; Yakobson,~B.~I. Two-dimensional mono-elemental semiconductor with electronically inactive defects: the case of phosphorus. \emph{Nano Lett.}
  \textbf{2014}, \emph{14}, 6782-6786\relax
\mciteBstWouldAddEndPuncttrue
\mciteSetBstMidEndSepPunct{\mcitedefaultmidpunct}
{\mcitedefaultendpunct}{\mcitedefaultseppunct}\relax
\EndOfBibitem
\bibitem[Liang et~al.(2014)Liang, Wang, Lin, Sumpter, Meunier, and Pan]{Liang}
Liang,~L.; Wang,~J.; Lin,~W.; Sumpter,~B.~G.; Meunier,~V.; Pan,~M. Electronic bandgap and edge reconstruction in phosphorene materials. \emph{Nano
  Lett.} \textbf{2014}, \emph{14}, 6400-6406\relax
\mciteBstWouldAddEndPuncttrue
\mciteSetBstMidEndSepPunct{\mcitedefaultmidpunct}
{\mcitedefaultendpunct}{\mcitedefaultseppunct}\relax
\EndOfBibitem
\bibitem[Mehboudi et~al.(2015)Mehboudi, Utt, Terrones, Harriss,
  Pacheco-Sanjuan, and Barraza-Lopez]{us5}
Mehboudi,~M.; Utt,~K.; Terrones,~H.; Harriss,~E.~O.; Pacheco-Sanjuan,~A.~A.;
  Barraza-Lopez,~S. Strain and the optoelectronic properties of nonplanar phosphorene monolayers. \emph{Proc. Natl. Acad. Sci. (USA)} \textbf{2015},
  \emph{112}, 5888-5892\relax
\mciteBstWouldAddEndPuncttrue
\mciteSetBstMidEndSepPunct{\mcitedefaultmidpunct}
{\mcitedefaultendpunct}{\mcitedefaultseppunct}\relax
\EndOfBibitem
\bibitem[Li et~al.(2015)Li, Guo, Cao, Liu, Lau, and Liu]{Liu}
Li,~X.-B.; Guo,~P.; Cao,~T.-F.; Liu,~H.; Lau,~W.-M.; Liu,~L.-M. Structures, stabilities, and electronic properties of defects in monolayer black phosphorus. \emph{Sci.
  Rep.} \textbf{2015}, \emph{5}, 10848\relax
\mciteBstWouldAddEndPuncttrue
\mciteSetBstMidEndSepPunct{\mcitedefaultmidpunct}
{\mcitedefaultendpunct}{\mcitedefaultseppunct}\relax
\EndOfBibitem
\bibitem[Nelson(2002)]{Nelson}
Nelson,~D.~R. \emph{Defects and Geometry in Condensed Matter Physics}, 1st ed.;
  Cambridge U. Press: Cambridge, UK, 2002\relax
\mciteBstWouldAddEndPuncttrue
\mciteSetBstMidEndSepPunct{\mcitedefaultmidpunct}
{\mcitedefaultendpunct}{\mcitedefaultseppunct}\relax
\EndOfBibitem
\bibitem[Vitelli et~al.(2006)Vitelli, Lucks, and Nelson]{vitelli}
Vitelli,~V.; Lucks,~J.; Nelson,~D. Crystallography on curved surfaces. \emph{Proc. Natl. Acad. Sci. (USA)}
  \textbf{2006}, \emph{103}, 12323-12328\relax
\mciteBstWouldAddEndPuncttrue
\mciteSetBstMidEndSepPunct{\mcitedefaultmidpunct}
{\mcitedefaultendpunct}{\mcitedefaultseppunct}\relax
\EndOfBibitem
\bibitem[Wales(2003)]{EL}
Wales,~D. \emph{Energy Landscapes}, 1st ed.; Cambridge U. Press: Cambridge,
  U.K., 2003\relax
\mciteBstWouldAddEndPuncttrue
\mciteSetBstMidEndSepPunct{\mcitedefaultmidpunct}
{\mcitedefaultendpunct}{\mcitedefaultseppunct}\relax
\EndOfBibitem
\bibitem[Vernizzi et~al.(2011)Vernizzi, Sknepnek, and {Olvera~de~la~Cruz}]{m1}
Vernizzi,~G.; Sknepnek,~R.; {Olvera~de~la~Cruz},~M. Platonic and Archimedean geometries in multicomponent elastic membranes. \emph{Proc. Natl. Acad.
  Sci. (USA)} \textbf{2011}, \emph{108}, 4292-4296\relax
\mciteBstWouldAddEndPuncttrue
\mciteSetBstMidEndSepPunct{\mcitedefaultmidpunct}
{\mcitedefaultendpunct}{\mcitedefaultseppunct}\relax
\EndOfBibitem
\bibitem[Sing et~al.(2014)Sing, Zwanikken, and {Olvera~de~la~Cruz}]{m2}
Sing,~C.; Zwanikken,~J.; {Olvera~de~la~Cruz},~M. Electrostatic control of block copolymer morphology. \emph{Nature Mater.}
  \textbf{2014}, \emph{13}, 694-698\relax
\mciteBstWouldAddEndPunctfalse
\mciteSetBstMidEndSepPunct{\mcitedefaultmidpunct}
{}{\mcitedefaultseppunct}\relax
\EndOfBibitem
\bibitem[Sussman et~al.(2015)Sussman, Cho, Castle, Gong, Jung, Yang, and
  Kamien]{Kamien3}
Sussman,~D.; Cho,~Y.; Castle,~T.; Gong,~X.; Jung,~E.; Yang,~S.; Kamien,~R. Algorithmic lattice kirigami: A route to pluripotent materials. \emph{Proc. Natl. Acad. Sci. (USA)}, \textbf{2015}, \emph{112}, 7449-7453\relax
\mciteBstWouldAddEndPuncttrue
\mciteSetBstMidEndSepPunct{\mcitedefaultmidpunct}
{\mcitedefaultendpunct}{\mcitedefaultseppunct}\relax
\EndOfBibitem
\bibitem[Lord et~al.(2006)Lord, Mackay, and Ranganathan]{ngnm}
Lord,~E.; Mackay,~A.; Ranganathan,~S. \emph{New Geometries for New Materials},
  1st ed.; Cambridge U. Press: Cambridge, U.K., 2006\relax
\mciteBstWouldAddEndPuncttrue
\mciteSetBstMidEndSepPunct{\mcitedefaultmidpunct}
{\mcitedefaultendpunct}{\mcitedefaultseppunct}\relax
\EndOfBibitem
\bibitem[Kamien(2002)]{Kamien1}
Kamien,~R. The geometry of soft materials: a primer. \emph{Rev. Mod. Phys.} \textbf{2002}, \emph{74}, 953-971\relax
\mciteBstWouldAddEndPuncttrue
\mciteSetBstMidEndSepPunct{\mcitedefaultmidpunct}
{\mcitedefaultendpunct}{\mcitedefaultseppunct}\relax
\EndOfBibitem
\bibitem[Castle et~al.(2014)Castle, Cho, Gong, Jung, Sussman, Yang, and
  Kamien]{Kamien2}
Castle,~T.; Cho,~Y.; Gong,~X.; Jung,~E.; Sussman,~D.; Yang,~S.; Kamien,~R. Making the cut: lattice kirigami rules. \emph{Phys. Rev. Lett.} \textbf{2014}, \emph{113}, 245502\relax
\mciteBstWouldAddEndPuncttrue
\mciteSetBstMidEndSepPunct{\mcitedefaultmidpunct}
{\mcitedefaultendpunct}{\mcitedefaultseppunct}\relax
\EndOfBibitem
\bibitem[Gonz{\'a}lez et~al.(1992)Gonz{\'a}lez, Guinea, and
  Vozmediano]{vozmediano92}
Gonz{\'a}lez,~J.; Guinea,~F.; Vozmediano,~M. Continuum approximation to fullerene molecules. \emph{Phys. Rev. Lett.}
  \textbf{1992}, \emph{69}, 172-175\relax
\mciteBstWouldAddEndPuncttrue
\mciteSetBstMidEndSepPunct{\mcitedefaultmidpunct}
{\mcitedefaultendpunct}{\mcitedefaultseppunct}\relax
\EndOfBibitem
\bibitem[Haddon(1993)]{Haddon}
Haddon,~R. Chemistry of the fullerenes: the manifestation of strain in a class of continuous aromatic molecules. \emph{Science} \textbf{1993}, \emph{261}, 1545-1550\relax
\mciteBstWouldAddEndPuncttrue
\mciteSetBstMidEndSepPunct{\mcitedefaultmidpunct}
{\mcitedefaultendpunct}{\mcitedefaultseppunct}\relax
\EndOfBibitem
\bibitem[Vozmediano et~al.(2010)Vozmediano, Katsnelson, and Guinea]{vozmediano}
Vozmediano,~M.; Katsnelson,~M.; Guinea,~F. Gauge fields in graphene. \emph{Phys. Rep.} \textbf{2010},
  \emph{496}, 109-148\relax
\mciteBstWouldAddEndPuncttrue
\mciteSetBstMidEndSepPunct{\mcitedefaultmidpunct}
{\mcitedefaultendpunct}{\mcitedefaultseppunct}\relax
\EndOfBibitem
\bibitem[Serrano-Andr{\'e}s et~al.(2006)Serrano-Andr{\'e}s, Merch{\'a}n, and
  Borin]{serrano}
Serrano-Andr{\'e}s,~L.; Merch{\'a}n,~M.; Borin,~A. Adenine and 2-aminopurine: paradigms of modern theoretical photochemistry. \emph{Proc. Natl. Acad. Sci.
  (USA)} \textbf{2006}, \emph{103}, 8691-8696\relax
\mciteBstWouldAddEndPuncttrue
\mciteSetBstMidEndSepPunct{\mcitedefaultmidpunct}
{\mcitedefaultendpunct}{\mcitedefaultseppunct}\relax
\EndOfBibitem
\bibitem[Ziletti et~al.(2015)Ziletti, Carvalho, Campbell, Coker, and
  {Castro~Neto}]{oxygenplanar1}
Ziletti,~A.; Carvalho,~A.; Campbell,~D.~K.; Coker,~D.~F.; {Castro~Neto},~A.~H. Oxygen defects in phosphorene. \emph{Phys. Rev. Lett.} \textbf{2015}, \emph{114}, 046801\relax
\mciteBstWouldAddEndPuncttrue
\mciteSetBstMidEndSepPunct{\mcitedefaultmidpunct}
{\mcitedefaultendpunct}{\mcitedefaultseppunct}\relax
\EndOfBibitem
\bibitem[Wang et~al.(2015)Wang, Pandey, and Karna]{oxygenplanar2}
Wang,~G.; Pandey,~R.; Karna,~S.~P. Phosphorene oxide: stability and electronic properties of a novel two-dimensional material. \emph{Nanoscale} \textbf{2015}, \emph{7}, 524-531\relax
\mciteBstWouldAddEndPuncttrue
\mciteSetBstMidEndSepPunct{\mcitedefaultmidpunct}
{\mcitedefaultendpunct}{\mcitedefaultseppunct}\relax
\EndOfBibitem
\bibitem[Yuan et~al.(2015)Yuan, Rudenko, and Katsnelson]{oxygenplanar3}
Yuan,~S.; Rudenko,~A.~N.; Katsnelson,~M.~I. Transport and optical properties of single- and bilayer black phosphorus with defects. \emph{Phys. Rev. B} \textbf{2015},
  \emph{91}, 115436\relax
\mciteBstWouldAddEndPuncttrue
\mciteSetBstMidEndSepPunct{\mcitedefaultmidpunct}
{\mcitedefaultendpunct}{\mcitedefaultseppunct}\relax
\EndOfBibitem
\bibitem[Wang et~al.(2015)Wang, Pandey, and Karna]{oxygenplanar4}
Wang,~G.; Pandey,~R.; Karna,~S.~P. Effects of extrinsic point defects in phosphorene: B, C, N, O, and F adatoms. \emph{Appl. Phys. Lett.} \textbf{2015},
  \emph{106}, 173104\relax
\mciteBstWouldAddEndPuncttrue
\mciteSetBstMidEndSepPunct{\mcitedefaultmidpunct}
{\mcitedefaultendpunct}{\mcitedefaultseppunct}\relax
\EndOfBibitem
\bibitem[Ziletti et~al.(2015)Ziletti, Carvalho, Trevisanutto, Campbell, Coker,
  and {Castro~Neto}]{oxygenplanar5}
Ziletti,~A.; Carvalho,~A.; Trevisanutto,~P.~E.; Campbell,~D.~K.; Coker,~D.~F.;
  {Castro~Neto},~A.~H. Phosphorene oxides: bandgap engineering of phosphorene by oxidation. \emph{Phys. Rev. B} \textbf{2015}, \emph{91},
  085407\relax
\mciteBstWouldAddEndPuncttrue
\mciteSetBstMidEndSepPunct{\mcitedefaultmidpunct}
{\mcitedefaultendpunct}{\mcitedefaultseppunct}\relax
\EndOfBibitem
\bibitem[Boukhvalov et~al.(2015)Boukhvalov, Rudenko, Prishchenko, Mazurenko,
  and Katsnelson]{oxygenplanar6}
Boukhvalov,~D.~W.; Rudenko,~A.~N.; Prishchenko,~D.~A.; Mazurenko,~V.~G.;
  Katsnelson,~M.~I.  Chemical modifications and stability of phosphorene with impurities: a first principles study. \emph{Phys. Chem. Chem. Phys.} \textbf{2015}, \emph{17}, 15209-15217\relax
\mciteBstWouldAddEndPuncttrue
\mciteSetBstMidEndSepPunct{\mcitedefaultmidpunct}
{\mcitedefaultendpunct}{\mcitedefaultseppunct}\relax
\EndOfBibitem
\bibitem[Sloan et~al.(2013)Sloan, Pacheco-Sanjuan, Wang, Horvath, and
  Barraza-Lopez]{us1}
Sloan,~J.~V.; Pacheco-Sanjuan,~A.~A.; Wang,~Z.; Horvath,~C.~M.;
  Barraza-Lopez,~S. Strain gauge fields for rippled graphene membranes under central mechanical load: An approach beyond first-order continuum elasticity. \emph{Phys. Rev. B} \textbf{2013}, \emph{87}, 155436\relax
\mciteBstWouldAddEndPuncttrue
\mciteSetBstMidEndSepPunct{\mcitedefaultmidpunct}
{\mcitedefaultendpunct}{\mcitedefaultseppunct}\relax
\EndOfBibitem
\bibitem[Barraza-Lopez et~al.(2013)Barraza-Lopez, Pacheco-Sanjuan, Wang, and
  Vanevi{\'c}]{us2}
Barraza-Lopez,~S.; Pacheco-Sanjuan,~A.~A.; Wang,~Z.; Vanevi{\'c},~M. Strain-engineering of graphene's electronic structure beyond continuum elasticity. \emph{Solid State Comm.} \textbf{2013}, \emph{166}, 70-75\relax
\mciteBstWouldAddEndPuncttrue
\mciteSetBstMidEndSepPunct{\mcitedefaultmidpunct}
{\mcitedefaultendpunct}{\mcitedefaultseppunct}\relax
\EndOfBibitem
\bibitem[Pacheco-Sanjuan et~al.(2014)Pacheco-Sanjuan, Wang, {Pour~Imani},
  Vanevic, and Barraza-Lopez]{us3}
Pacheco-Sanjuan,~A.~A.; Wang,~Z.; {Pour~Imani},~H.; Vanevic,~M.;
  Barraza-Lopez,~S. Graphene's morphology and electronic properties from discrete differential geometry. \emph{Phys. Rev. B} \textbf{2014}, \emph{89},
  121403(R)\relax
\mciteBstWouldAddEndPuncttrue
\mciteSetBstMidEndSepPunct{\mcitedefaultmidpunct}
{\mcitedefaultendpunct}{\mcitedefaultseppunct}\relax
\EndOfBibitem
\bibitem[Pacheco-Sanjuan et~al.(2014)Pacheco-Sanjuan, Mehboudi, Harriss,
  Terrones, and Barraza-Lopez]{us4}
Pacheco-Sanjuan,~A.~A.; Mehboudi,~M.; Harriss,~E.~O.; Terrones,~H.;
  Barraza-Lopez,~S. Quantitative chemistry and the discrete geometry of conformal atom-thin crystals. \emph{ACS Nano} \textbf{2014}, \emph{8}, 1136–1146\relax
\mciteBstWouldAddEndPuncttrue
\mciteSetBstMidEndSepPunct{\mcitedefaultmidpunct}
{\mcitedefaultendpunct}{\mcitedefaultseppunct}\relax
\EndOfBibitem
\bibitem[Sorkin and Zhang(2015)Sorkin, and Zhang]{edges2}
Sorkin,~V.; Zhang,~Y.~W. The structure and elastic properties of phosphorene edges. \emph{Nanotechnology} \textbf{2015}, \emph{26},
  235707\relax
\mciteBstWouldAddEndPuncttrue
\mciteSetBstMidEndSepPunct{\mcitedefaultmidpunct}
{\mcitedefaultendpunct}{\mcitedefaultseppunct}\relax
\EndOfBibitem
\bibitem[Zhu and Tom{\'a}nek(2014)Zhu, and Tom{\'a}nek]{Tomanek1}
Zhu,~Z.; Tom{\'a}nek,~D. Semiconducting layered blue phosphorus: a computational study. \emph{Phys. Rev. Lett.} \textbf{2014}, \emph{112},
  176802\relax
\mciteBstWouldAddEndPuncttrue
\mciteSetBstMidEndSepPunct{\mcitedefaultmidpunct}
{\mcitedefaultendpunct}{\mcitedefaultseppunct}\relax
\EndOfBibitem
\bibitem[Guan et~al.(2014)Guan, Zhu, and Tom{\'a}nek]{Tomanek2}
Guan,~J.; Zhu,~Z.; Tom{\'a}nek,~D. Phase coexistence and metal-insulator transition in few-layer phosphorene: a computational study. \emph{Phys. Rev. Lett.} \textbf{2014},
  \emph{113}, 046804\relax
\mciteBstWouldAddEndPuncttrue
\mciteSetBstMidEndSepPunct{\mcitedefaultmidpunct}
{\mcitedefaultendpunct}{\mcitedefaultseppunct}\relax
\EndOfBibitem
\bibitem[Guan et~al.(2014)Guan, Zhu, and Tom{\'a}nek]{Tomanek3}
Guan,~J.; Zhu,~Z.; Tom{\'a}nek,~D. Tiling phosphorene. \emph{ACS Nano} \textbf{2014}, \emph{8},
  12763–12768\relax
\mciteBstWouldAddEndPuncttrue
\mciteSetBstMidEndSepPunct{\mcitedefaultmidpunct}
{\mcitedefaultendpunct}{\mcitedefaultseppunct}\relax
\EndOfBibitem
\bibitem[Wu et~al.(2015)Wu, Fu, Zhou, Yao, and Zeng]{Wu}
Wu,~M.; Fu,~H.; Zhou,~L.; Yao,~K.; Zeng,~X. Nine new phosphorene polymorphs with non-honeycomb structures: a much extended family. \emph{Nano Lett.} \textbf{2015},
  \emph{15}, 3557-3562\relax
\mciteBstWouldAddEndPuncttrue
\mciteSetBstMidEndSepPunct{\mcitedefaultmidpunct}
{\mcitedefaultendpunct}{\mcitedefaultseppunct}\relax
\EndOfBibitem
\bibitem[Car and Parinello(1985)Car, and Parinello]{Car}
Car,~R.; Parinello,~M. Unified approach for molecular dynamics and density-functional theory. \emph{Phys. Rev. Lett.} \textbf{1985}, \emph{55},
  2471-2474\relax
\mciteBstWouldAddEndPuncttrue
\mciteSetBstMidEndSepPunct{\mcitedefaultmidpunct}
{\mcitedefaultendpunct}{\mcitedefaultseppunct}\relax
\EndOfBibitem
\bibitem[Banhart et~al.(2011)Banhart, Kotakoski, and Krasheninnikov]{Kotakoski}
Banhart,~F.; Kotakoski,~J.; Krasheninnikov,~A.~V. Structural defects in graphene. \emph{ACS Nano}
  \textbf{2011}, \emph{5}, 26-41\relax
\mciteBstWouldAddEndPuncttrue
\mciteSetBstMidEndSepPunct{\mcitedefaultmidpunct}
{\mcitedefaultendpunct}{\mcitedefaultseppunct}\relax
\EndOfBibitem
\bibitem[He et~al.(2010)He, Koepke, Barraza-Lopez, and Lyding]{NanoLetters}
He,~K.~T.; Koepke,~J.~C.; Barraza-Lopez,~S.; Lyding,~J.~W. Separation-dependent electronic transparency of monolayer graphene membranes on III-V semiconductor substrates. \emph{Nano Lett.}
  \textbf{2010}, \emph{10}, 3446-3452\relax
\mciteBstWouldAddEndPuncttrue
\mciteSetBstMidEndSepPunct{\mcitedefaultmidpunct}
{\mcitedefaultendpunct}{\mcitedefaultseppunct}\relax
\EndOfBibitem
\bibitem[Soler et~al.(2002)Soler, Artacho, Gale, Garc{\'\i}a, Junquera,
  Ordej{\'o}n, and S{\'a}nchez-Portal]{SIESTA1}
Soler,~J.~M.; Artacho,~E.; Gale,~J.~D.; Garc{\'\i}a,~A.; Junquera,~J.;
  Ordej{\'o}n,~P.; S{\'a}nchez-Portal,~D. The SIESTA method for {\em ab initio} order-N materials simulation. \emph{J. Phys.: Condens. Matter}
  \textbf{2002}, \emph{14}, 2745\relax
\mciteBstWouldAddEndPuncttrue
\mciteSetBstMidEndSepPunct{\mcitedefaultmidpunct}
{\mcitedefaultendpunct}{\mcitedefaultseppunct}\relax
\EndOfBibitem
\bibitem[Artacho et~al.(2008)Artacho, Anglada, Di{\'e}guez, Gale, Garc{\'\i}a,
  Junquera, Martin, Ordej{\'o}n, Pruneda, S{\'a}nchez-Portal, and
  Soler]{SIESTA2}
Artacho,~E.; Anglada,~E.; Di{\'e}guez,~O.; Gale,~J.~D.; Garc{\'\i}a,~A.;
  Junquera,~J.; Martin,~R.~M.; Ordej{\'o}n,~P.; Pruneda,~J.~M.;
  S{\'a}nchez-Portal,~D.; Soler,~J.~M. The SIESTA method; developments and applicability. \emph{J. Phys.: Condens. Matter}
  \textbf{2008}, \emph{20}, 064208\relax
\mciteBstWouldAddEndPuncttrue
\mciteSetBstMidEndSepPunct{\mcitedefaultmidpunct}
{\mcitedefaultendpunct}{\mcitedefaultseppunct}\relax
\EndOfBibitem
\bibitem[Weischedel et~al.(2012)Weischedel, Tuganov, Hermansson, Linn, and
  Wardetzky]{conf}
Weischedel,~C.; Tuganov,~A.; Hermansson,~T.; Linn,~J.; Wardetzky,~M. (2012)
  Construction of discrete shell models by geometric finite differences, The 2nd Joint Conference on Multibody System Dynamics, May 29--June 1, 2012, Stuttgart, Germany\relax
\mciteBstWouldAddEndPuncttrue
\mciteSetBstMidEndSepPunct{\mcitedefaultmidpunct}
{\mcitedefaultendpunct}{\mcitedefaultseppunct}\relax
\EndOfBibitem
\bibitem[Bobenko et~al.(2008)Bobenko, Schr{\"o}der, Sullivan, and Ziegler]{DDG}
Bobenko,~A., Schr{\"o}der,~P., Sullivan,~J., Ziegler,~G., Eds. \emph{Discrete
  Differential Geometry}, 1st ed.; Oberwolfach Seminars; Birkh{\"a}user: Basel,
  Switzerland, 2008\relax
\mciteBstWouldAddEndPuncttrue
\mciteSetBstMidEndSepPunct{\mcitedefaultmidpunct}
{\mcitedefaultendpunct}{\mcitedefaultseppunct}\relax
\EndOfBibitem
\end{mcitethebibliography}

\providecommand*\mcitethebibliography{\thebibliography}
\csname @ifundefined\endcsname{endmcitethebibliography}
  {\let\endmcitethebibliography\endthebibliography}{}

\end{document}